\documentclass[aps,pra,twocolumn]{revtex4-1}
\usepackage{amsmath}
\usepackage{graphicx}
\usepackage{mathrsfs}

\newcommand{\be}{\begin{equation}}
\newcommand{\ee}{\end{equation}}

\begin{document}

\title{Interplay between boundaries and disorder in coupled optical waveguides}

\author{Clinton Thompson}
\affiliation{Department of Physics, 
Indiana University Purdue University Indianapolis (IUPUI), 
Indianapolis, Indiana 46202, USA}

\author{Yogesh N. Joglekar}
\affiliation{Department of Physics, 
Indiana University Purdue University Indianapolis (IUPUI), 
Indianapolis, Indiana 46202, USA}

\author{Gautam Vemuri}
\affiliation{Department of Physics, 
Indiana University Purdue University Indianapolis (IUPUI), 
Indianapolis, Indiana 46202, USA}
\date{\today}

\begin{abstract}
We investigate the effects of weak disorder on the time evolution of a wave packet in an array of optical waveguides with parity-symmetric evanescent coupling and, open or periodic boundary conditions. For an open array, when the disorder is unable to suppress the ballistic expansion, we find that the light partially localizes to its initial waveguide and the parity-symmetric waveguide when a single waveguide is excited. For an array with periodic boundary condition, the light localizes in the initial waveguide and its antipodal waveguide. Through a model where the boundaries of the array are only partially reflective, we quantify and investigate the continuous crossover between these two regimes. Our results show that disorder-induced localization in finite arrays with very weak disorder is strongly affected by the boundary effects that have been hitherto ignored. 
\end{abstract}

\maketitle


\section{Introduction}
\label{sec:Intro}
In recent years, arrays of evanescently coupled waveguides in two- and three-dimensions have become a paradigm for the realization of one and two-dimensional tight-binding Hamiltonians respectively. Historically, light and matter have been considered two different entities. Thus, the ability to use electromagnetic waves in dielectrics to faithfully simulate a tight-binding Hamiltonian, primarily used in condensed matter, is a result of the advent of technology that is used to fabricate the waveguides. Some of the remarkable phenomena that have been experimentally observed in waveguide arrays are Anderson localization~\cite{Lahini2008}, Bloch oscillations~\cite{Peschel1998}, the Aharonov-Bohm effect~\cite{Fang2012}, quantum random walks~\cite{Perets2008}, and Zener tunneling~\cite{Trompeter2006}.  Although coupled waveguide arrays and tight-binding lattice models from condensed matter systems are mathematically identical at a single-particle level, there are some crucial differences. 

The first major advantage of using light in coupled-waveguide arrays as a realization of a quantum particle on a tight-binding lattice is that the light-intensity distribution, which corresponds to the ``quantum probability distribution'', is measured directly. In contrast, the probability distribution of a quantum particle hopping on a lattice is measured indirectly, typically through the conductance measurements~\cite{Gantmakher2005}. The second advantage is that the phenomena of interest - be it Anderson localization or Bloch oscillations - are observed over a lengthscale of a few centimeters in waveguide arrays instead of a few microns for electrons~\cite{Datta1997}. Adjusting the incident angle of the light coupled to a particular waveguide controls the diffraction of the light and allows for a diffractionless array~\cite{Eisenberg2000}. In addition, the photon-photon interaction in the dielectric, characterized by the third-order susceptibility $\chi^{(3)}$ is extremely small at relevant light intensities and the decoherence noise in a waveguide array is negligible~\cite{Segev2013}; in contrast, for electrons in a tight-binding lattice, Coulomb interaction is dominant, and to ensure that the decoherence-time from the lattice phonons is maximized, the system must be cooled to temperatures much below the Debye temperature $T_D$~\cite{Gantmakher2005}. The third major advantage of using coupled waveguide arrays is that one can sample the entire energy band of a tight-binding lattice using photons, as a photon injected into a single waveguide - localized to a single lattice site - is a linear superposition of all energy eigenstates with equal weights. In contrast, for condensed-matter systems, only the electrons with energies within $k_BT$ of the Fermi energy are sampled where $k_B$ is the Boltzmann constant and $T$ is the absolute temperature~\cite{DiVentra2008}. Lastly, coupled waveguide arrays allow systematic investigation of finite size and boundary-condition effects, because the typical number of waveguides in an array is small, $N\lesssim 100$, whereas for condensed-matter lattices, the number of lattice sites is $N>10^9$~\cite{Ashcroft1976}. 

Due to these factors, Anderson localization - a quintessential single-particle phenomenon in condensed matter systems - has been extensively studied via disordered waveguide arrays~\cite{Segev2013}. These studies, using bulk lasers or single-photon sources, have been carried out in ``large arrays'' with sufficiently strong disorder to ensure that the partial waves returning from the array ends do not significantly contribute to the non-ballistic, localized intensity component. In such large arrays,  ``disorder averaging'' is carried out by simply using different initial waveguides to input the light, thus obviating the need for a large number of samples. On the other hand, the tremendous versatility of coupled waveguide arrays means that, in contrast to the lattices in condensed-matter systems,  ``Anderson localization'' can be investigated in small arrays with tunable position-dependent hopping, where the effects of boundary conditions are not negligible. The competition among ballistic expansion, disorder-induced diffusion, and reflection or transparency at the boundaries raises a number of questions that are relevant for small, finite lattices, but are not applicable in the thermodynamic limit. How does a wavepacket evolve in the presence of extremely weak disorder, such that the time required to reach a steady state is longer than that to reach the array boundaries? How do different boundary conditions change the disorder-averaged, steady-state, localization intensity profile?  

In this paper, we numerically and analytically investigate these questions in waveguide arrays with constant or position-dependent, parity-symmetric coupling profiles. For an extremely weak disorder, we find that that, in addition to its initial waveguide, the light also localizes in either the parity-symmetric waveguide or  the antipodal waveguide in arrays with open or periodic boundary conditions respectively. The crossover between open and periodic boundary conditions  and its implications to localization are then discussed. We quantify boundary effect and show that it scales inversely with the lattice size. 


\section{Effect of a weak disorder}
\label{sec:weak}
The Hamiltonian describing the time-evolution of light inside the array of $N$ coupled, single-mode waveguides is given by 
\be 
\label{eq:T.B. Ham.}
H=\sum_{j=1}^N \beta_{j}a_{j}^{\dag }a_{j}+ \sum_{j=1}^{N-1} C_{\alpha}(j)( a_{j+1}^{\dag }a_{j}+a_{j}^{\dag }a_{j+1})+H_b,
\ee
where $\beta_{j}$ is the linear propagation constant, with units of frequency, for the $j^\mathrm{th}$ waveguide, $a_{j}^{\dag}(a_{j})$ is the creation (annihilation) operator for the single-mode electric-field in the $j^\mathrm{th}$ waveguide, and $C_{\alpha}(j)$ is the coupling constant or hopping amplitude between waveguides $j$ and $j+1$. The ``boundary'' Hamiltonian is given by $H_b=C_\alpha(N)( a^{\dagger}_N a_1+ a^\dagger_1 a_N)$ and allows us to tune between an open chain, specified by $C_\alpha(N)=0$, and a ring, specified by $C_\alpha(N)\neq 0$. (We have taken $\hbar=1$.) 

We emphasize at this point that a one-dimensional open chain of coupled optical waveguides is experimentally realized in a two-dimensional structure, where the second dimension denotes the waveguide length or, equivalently, the time. A ring, or an array with periodic boundary condition, however, can only be realized by using the ``boundary'' of a two-dimensional coupled waveguide array, where the waveguides run along the third dimension. Such three-dimensional structures, which can model the time-evolution of a quantum particle on two-dimensional lattices with different lattice structures, have been experimentally investigated. Thus, although one can transition from open to periodic boundary conditions in a unified model via $H_b$, experimentally, the two cases represent very different systems. 

To preserve the left-right symmetry in the lattice, we focus on parity-symmetric coupling functions of the form
\be 
\label{eq:tunnelingrate} 
C_{\alpha}(j)=C[j(N-j)]^{\alpha/2}=C_\alpha(N-j).
\ee  
Note that the prototypical constant-coupling case corresponds to $\alpha=0$. For $\alpha>0$, the coupling at the edges is smaller than that at the array center, whereas for $\alpha<0$, the coupling is maximum at the array edges. This means that the light prefers to be at the center of the array when $\alpha >0$ while for negative values of $\alpha$, the light prefers to be at the edges of the array. Waveguide arrays with this form of coupling function can be fabricated by symmetrically increasing (decreasing) the distance between adjacent waveguides for negative (positive) values of $\alpha$. They have been experimentally achieved for $\alpha=0$~\cite{Christodoulides2003} and $\alpha=1$~\cite{Bellec2012}. Even in the absence of on-site potentials $\beta_j$, the position-dependent coupling functions lead to remarkable wave packet dynamics that can be tuned by $\alpha$~\cite{Joglekar2011}. 

To study the wave packet evolution in the array, an initial site $J_0$ is chosen and the time-evolved wave function $|\psi(t) \rangle$, is calculated by applying the unitary time-evolution operator, $U(t)=\exp(-iHt)$, to the initial state $|\psi(0)\rangle=|j_0\rangle$. The site- and time-dependent intensity at waveguide $j$ as a function of time is therefore $I_{j}(t)=|\langle j|\psi (t) \rangle |^{2}$. We  use the inverse of the disorder-free bandwidth $\Delta_\alpha(N)$ as the characteristic time-scale for the system, $\tau_{\alpha}(N)=1/\Delta_{\alpha}(N)$. Eq.(\ref{eq:tunnelingrate}), and the fact that 
bandwidth of the hopping Hamiltonian is determined by its maximum hopping rate, imply that $\Delta_{\alpha}(N)\sim C N^{\alpha}$ for $\alpha > 0$ and $\Delta_{\alpha} \sim C N^{-|\alpha|/2}$ for $\alpha < 0$ \cite{Joglekar2011a}. Thus, in samples with a fixed waveguide length $\alpha>0$ allows exploration of long time-scales, whereas $\alpha<0$ allows the investigation of very short time-scales. We will be focused on the constant coupling case, $\alpha=0$, for most of this paper. 

The disorder in the Hamiltonian (\ref{eq:T.B. Ham.}) can either be introduced through the on-site potentials $\beta_{j}$ or through the waveguide couplings $C_\alpha(j)$. The former destroys the particle-hole symmetry of the resultant spectrum whereas the latter does not. The output intensity profiles are quantitatively similar for both cases, although the two-particle correlations are sensitive to the source of the disorder~\cite{Lahini2011}. We introduce disorder through the on-site potentials, $\beta_{j}=\beta+\delta\beta_j$ where $\delta\beta_j$ is the random disorder.  We use a disorder with zero mean and variance $\sigma^2$, and since the results are independent of the probability distribution, we use a Gaussian-distributed disorder. The number of disorder realizations $N_r\gg 1$ is varied to ensure that the disorder-averaged results are independent of $N_r$. Thus, the disorder-averaged site- and time-dependent intensity is defined as 
\be
\label{eq:do}
\langle I_j(t)\rangle= \frac{1}{N_r} \sum_{k=1}^{N_r} I_j(t,\sigma_k)
\ee
where $I_j(t,\sigma_k)$ denotes the intensity profile for the $k^\mathrm{th}$ disorder realization with the fixed disorder variance $\sigma$. 

\begin{figure*}[htpb]
\centering
\includegraphics[width=0.60\textwidth]{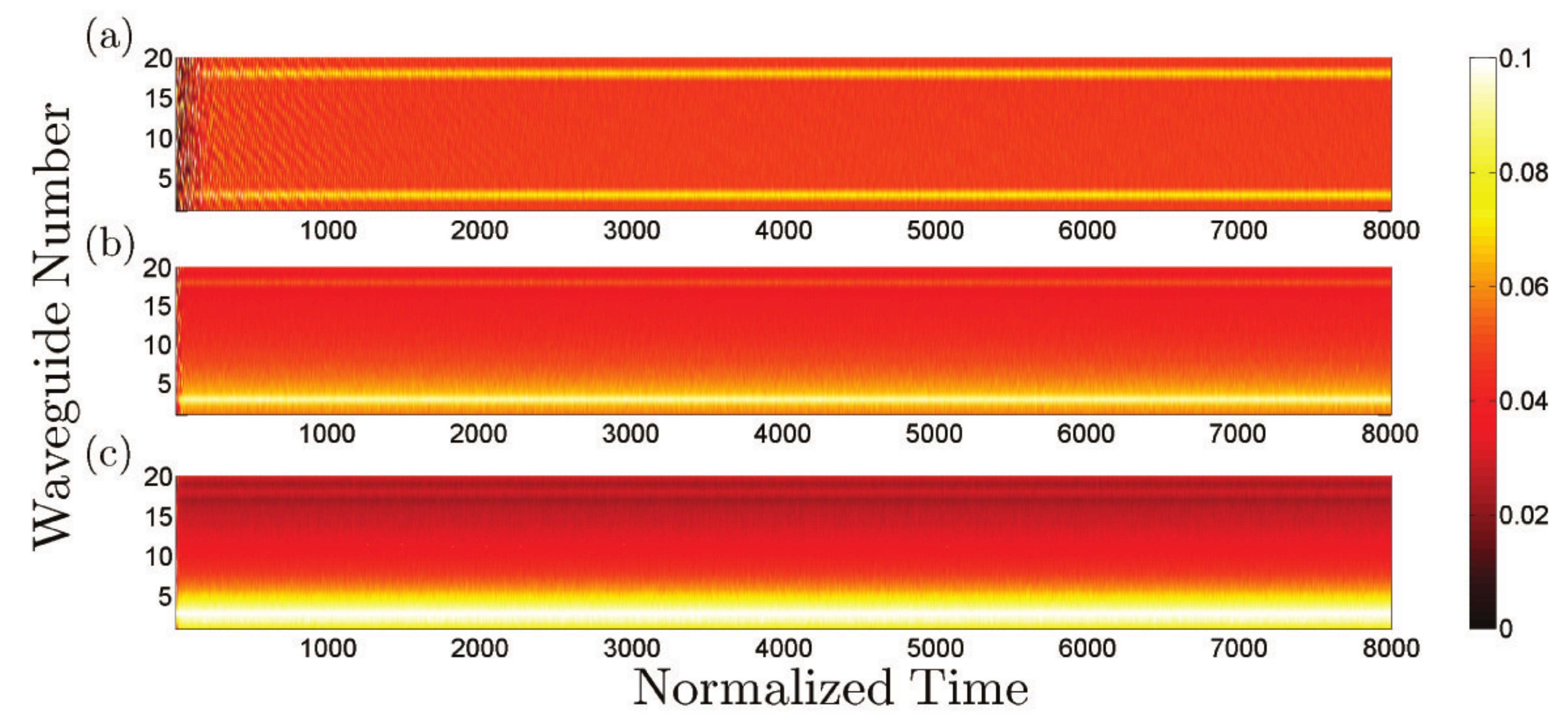}
\includegraphics[width=0.34\textwidth]{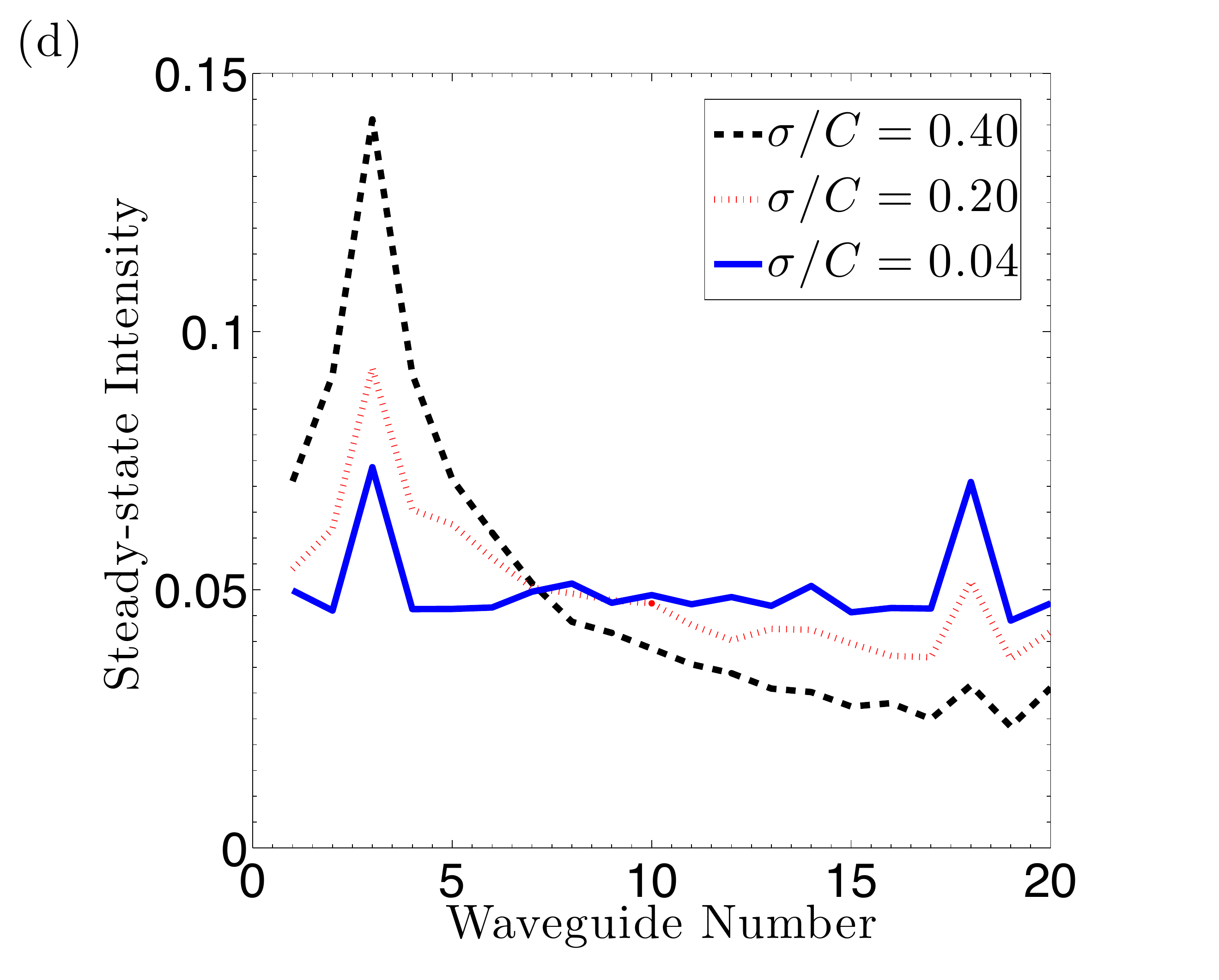}
\caption{(Color Online) Disorder-averaged, time- and site-dependent intensity for an array with $N =20$ waveguides and (a) $ \sigma /C =0.04$, (b) $ \sigma /C =0.20$, and (c) $ \sigma /C =0.40$. The initial state is localized at site $j_0=3$. Panel (d) shows the steady-state intensity profiles for  
$\sigma /C =0.40$ (black dashed line), $\sigma /C =0.2$ (red dotted line), and $\sigma /C =0.04$ (blue solid line). In addition to the expected localization in initial waveguide $j_0=3$, the light also localizes to its parity-symmetric waveguide $N+1-j_{0}=18$. As the disorder strength $\sigma$ increases, the amount of light that is localized at the parity-symmetric waveguide decreases.}
\label{fig1}
\end{figure*}
For a {\it finite} one-dimensional disordered lattice with open boundary conditions, $H_b=0$, the disorder-averaged intensity at the initial site depends upon the lattice size, the propagation time or distance along the waveguide, and the strength of the disorder. In contrast, for an infinite lattice, a vanishingly small, but nonzero, disorder exponentially localizes all eigenstates~\cite{Abrahams1979}. This exponential Anderson localization has been observed in large waveguide arrays, where the ballistic expansion time is larger than the time required to traverse the length of the waveguide~\cite{Perets2008}. When the initial state is localized at the center of such a large array, it effectively models an infinite array with no boundary effects, or when the initial waveguide is at the edge of the array, it models the semi-infinite case~\cite{Szameit2010}. The only (weak)  dependence of the localized fraction arises from proximity of the initially excited waveguide to the array  boundary; it takes larger disorder $\sigma\sim\Delta_\alpha$  for an input at the boundary to localize than if the initial position is near the center of the array for a given propagation time~\cite{Szameit2010}. Here we are interested in the localization dynamics when the disorder is weak, $\sigma \ll \Delta_{\alpha}^{(0)}$, and the light reaches or undergoes multiple reflections at the array boundaries.

Figures \ref{fig1}(a)-\ref{fig1}(c) show the disorder-averaged, time- and site-dependent intensity $\langle I_j(t)\rangle$ for an initial state $|j_0\rangle$ in an array with $N=20$ waveguides. This array has open boundary conditions, $C(N)=0$. Thus the average intensity per site is given by $1/N=0.05$. The hopping function is constant, $\alpha=0$, and the number of disorder realizations is $N_r=2000$. Panel (d) shows the steady-state intensity $\langle I_j\rangle$ for panels (a)-(c). When $\sigma/C=0.4$, panel (c), we see that the light is mostly localized to the initial waveguide, $j_0=3$ and the intensity profile decays monotonically on both sides of $j_0$, as shown in panel (d)  (black dashed line). This behavior is consistent with the expected localization, where the maximum value of the intensity on the initial site is thrice the average intensity.  When the disorder is lowered, $\sigma/C=0.2$, panel (b) shows the emergence of a clear local maximum at the parity-symmetric waveguide as well, and panel (d) shows that the steady-state intensity at the original waveguide is reduced (red dotted line). For a very weak disorder, $\sigma/C=0.04$, panel (a), we see a clear emergence of two-peaked structure in the steady-state intensity, and panel (d) shows that the steady-state weights at the original and parity-symmetric waveguides are approximately equal (blue solid line). Thus, we see that the exponentially decreasing localized, steady-state intensity profile at large disorder strength is replaced by a non-monotonic intensity profile that has two local maxima at $j_0$ and $N+1-j_0$. These two local maxima occur due to interference from multiple reflections at the array edges and localization due to the scattering from disorder potential, and therefore do not require a constant coupling function. 

\begin{figure}[htpb]
\centering
\includegraphics[width=\columnwidth]{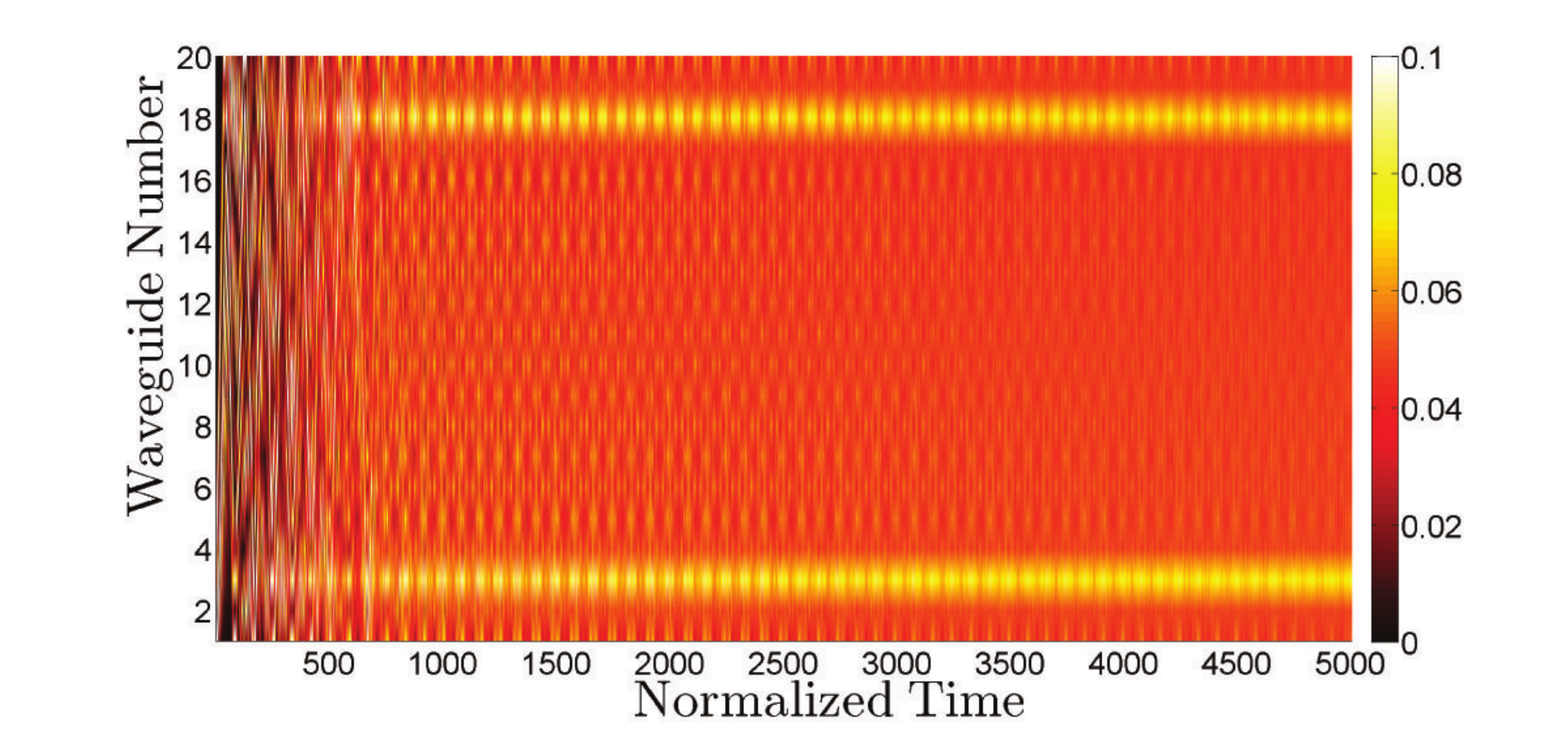}
\caption{(Color Online) $\langle I_j(t)\rangle$ in an $N=20$ array with coupling function $\alpha=-1$,  $\sigma/\Delta_\alpha =0.0125$, and an input at $j_0=3$. Time is normalized in units of $\tau_\alpha=1/\Delta_\alpha$ and here, $\Delta_\alpha\sim C/\sqrt{20}=0.023 C$. The light localizes at the initial waveguide and its parity-symmetric counterpart $N+1-j_0=18$. This two-channel localization is due to the interference from the array edges. }
\label{fig2}
\end{figure}
Figure~\ref{fig2} shows a representative localization in an $N=20$ open array with $\alpha=-1$, which means the coupling at the array ends is larger than that at the center of the array. The disorder-averaged intensity $\langle I_j(t)\rangle$ is obtained with $N_r=1000$ realizations of a weak disorder, $\sigma/\Delta_\alpha=0.0125$. We clearly see that $\langle I_j(t)\rangle$ has two local maxima at waveguides $j_0=3$ and $N+1-j_0=18$, and that the maximum intensity is twice the average intensity per site. This two-channel localization is observable in any array with parity-symmetric waveguide coupling when the disorder is weak, $\sigma/\Delta_{\alpha}\ll 1$~\cite{Thompson2012}. 

\begin{figure*}[t!]
\centering
\includegraphics[width=0.60\textwidth]{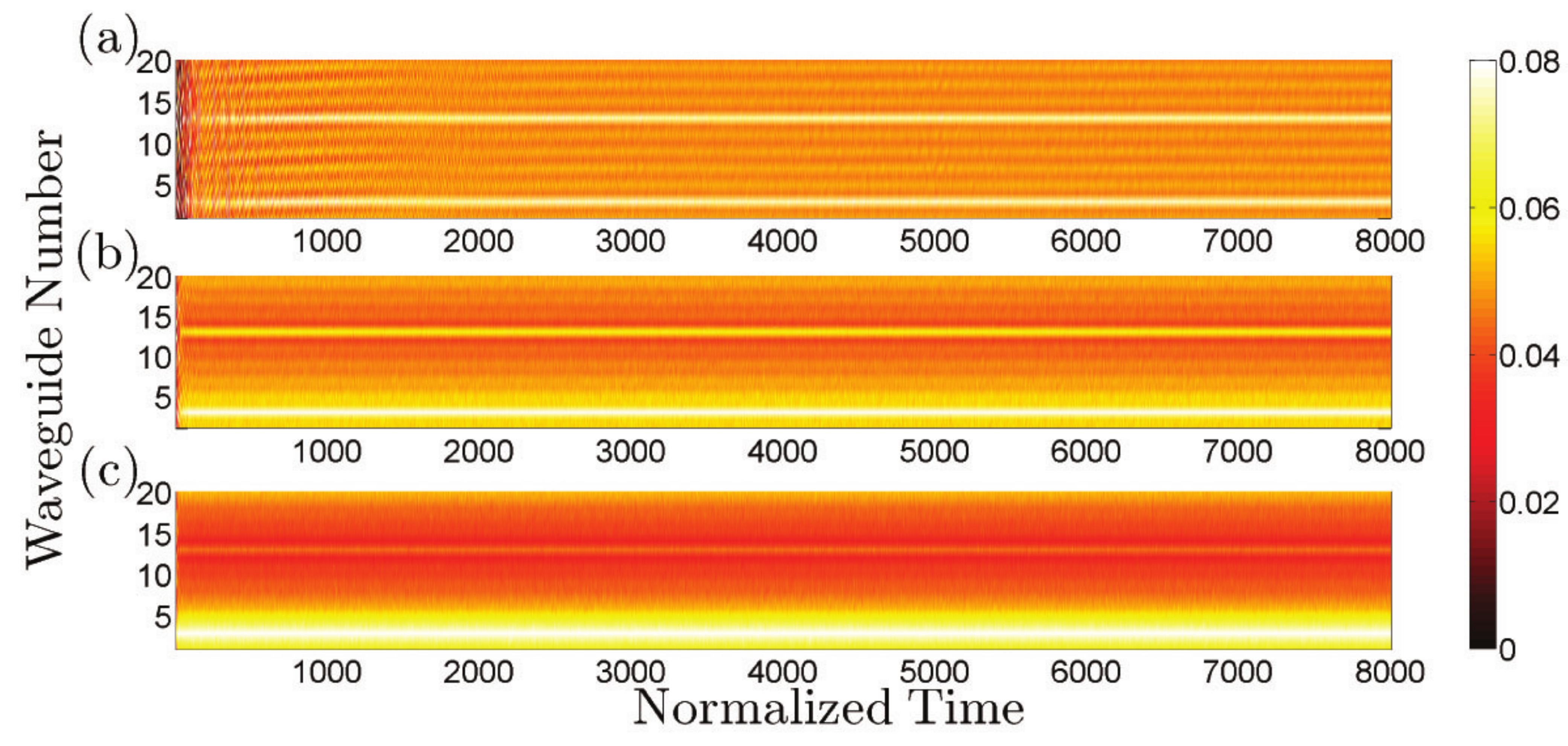}
\includegraphics[width=0.35\textwidth]{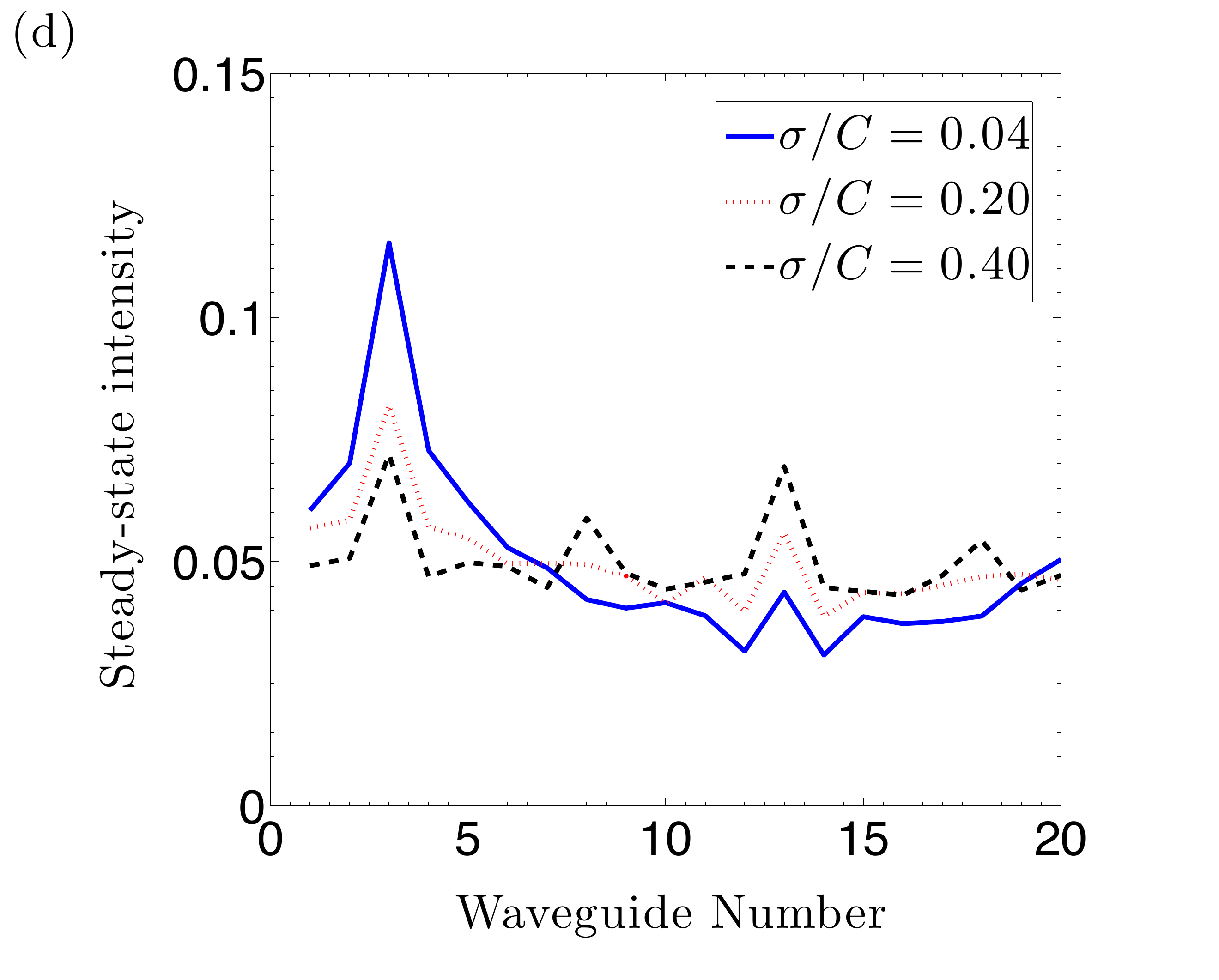}
\caption{(Color Online) Disorder-averaged intensity for a uniform, $N =20$ array with periodic boundary condition, and disorder strength (a) $\sigma /C =0.04$, (b) $ \sigma /C =0.20$, and (c) $ \sigma /C =0.40$. The light localizes in the initial waveguide $j_0=3$ and the antipodal waveguide $N/2+j_0=13$ as a result of constructive interference between the two paths joining them. Panel (d) shows the steady-state intensity profile $\langle I_j\rangle$ for $\sigma /C =0.04$ (blue solid line), $ \sigma /C =0.20$ (red dashed line), and $\sigma /C =0.40$ (black dotted line). As the disorder strength increases, the fraction of light intensity that localizes at the antipodal waveguide decreases.}
\label{fig4}
\end{figure*} 
To elucidate further the origin of the two-site localization phenomenon seen in small lattices in the presence of a weak disorder, we now consider a uniform array with periodic conditions, $H_b\neq 0$, with $C(N)=C$. We remind the reader that an experimental realization of a one-dimensional array with periodic boundary condition requires a two-dimensional waveguide lattice, with light propagation along the remaining, third direction. Figures ~\ref{fig4}(a)-\ref{fig4}(c) show the disorder-averaged intensity $\langle I_j(t)\rangle$ for an array with $N=20$ waveguides and the initial input in waveguide $j_0=3$. When the disorder is strong, $\sigma/C=0.4$, panel (c), light is localized into the input location, and the intensity decays monotonically on both sides of it, although there are hints of a local maximum at the antipodal position, $N/2+j_0$ (mod $N$)=13. As the disorder is reduced to $\sigma/C=0.20$, panel (b), and $\sigma/C=0.04$, panel (a), we find that the fraction of intensity localized to the initial waveguide $j_0$ reduces while that at the antipodal point $N/2+j_0$ (mod $N$) increases. These trends are quantified in panel (d) with the steady-state, disorder-averaged intensity $\langle I_j\rangle$, which shows that as the disorder strength decreases, the localized intensity weight shifts from the input location to the antipodal location. Note that since the overall disorder is weak compared to the bandwidth, $\sigma\ll 4C$, the maximum fraction localized to the initial site is just about twice the average intensity per site $1/N$. We emphasize that in this case, with periodic boundary conditions, the second peak arises from the competition between constructive interference at the antipodal point due to identical path lengths, and the localization due to scattering from on-site disorder. Therefore, {\it if the array consists of an odd number of waveguides, the second peak disappears}, as two paths that the light can traverse to any other waveguide have unequal path lengths.

\begin{figure*}[htpb]
\includegraphics[width=0.60\textwidth]{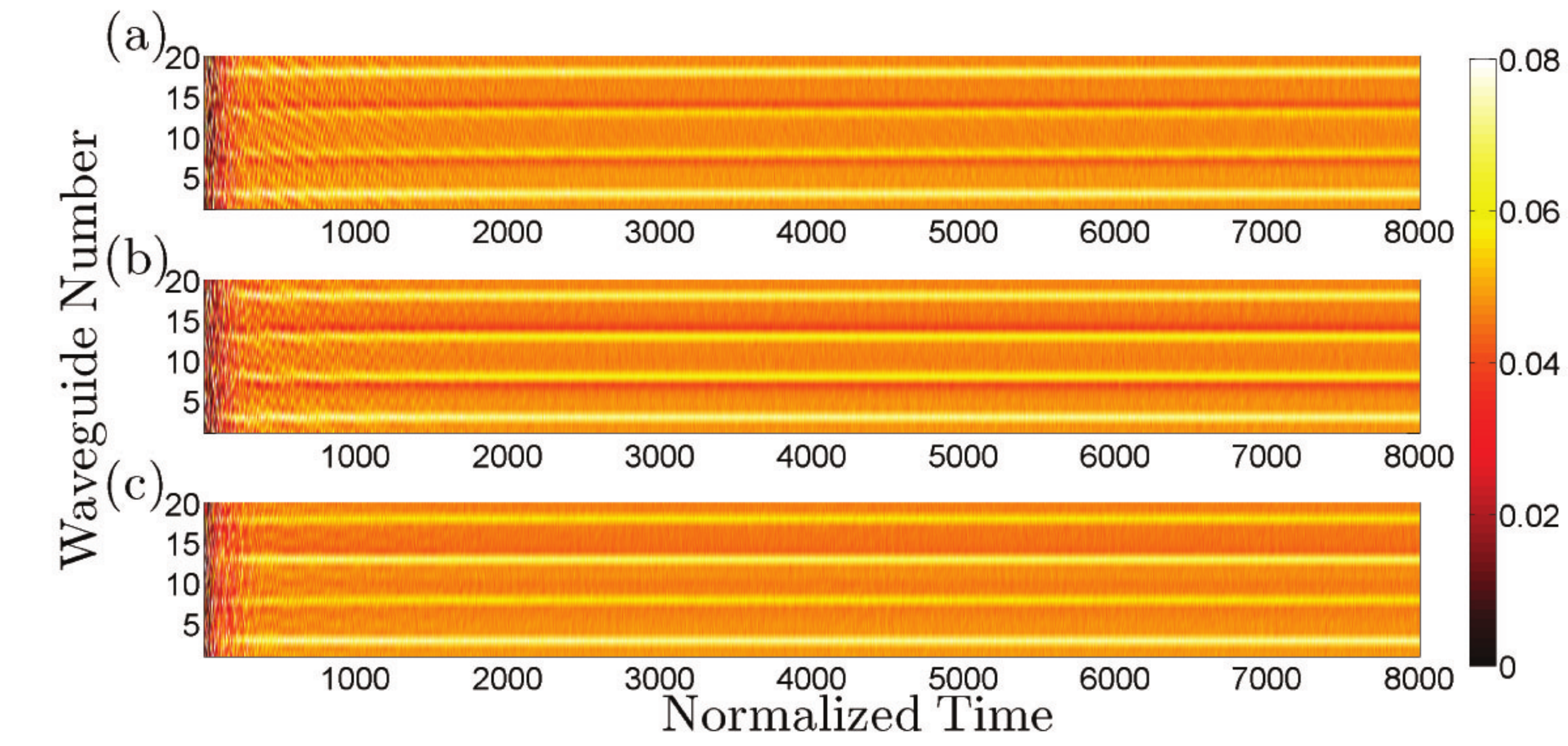}
\includegraphics[width=0.35\textwidth]{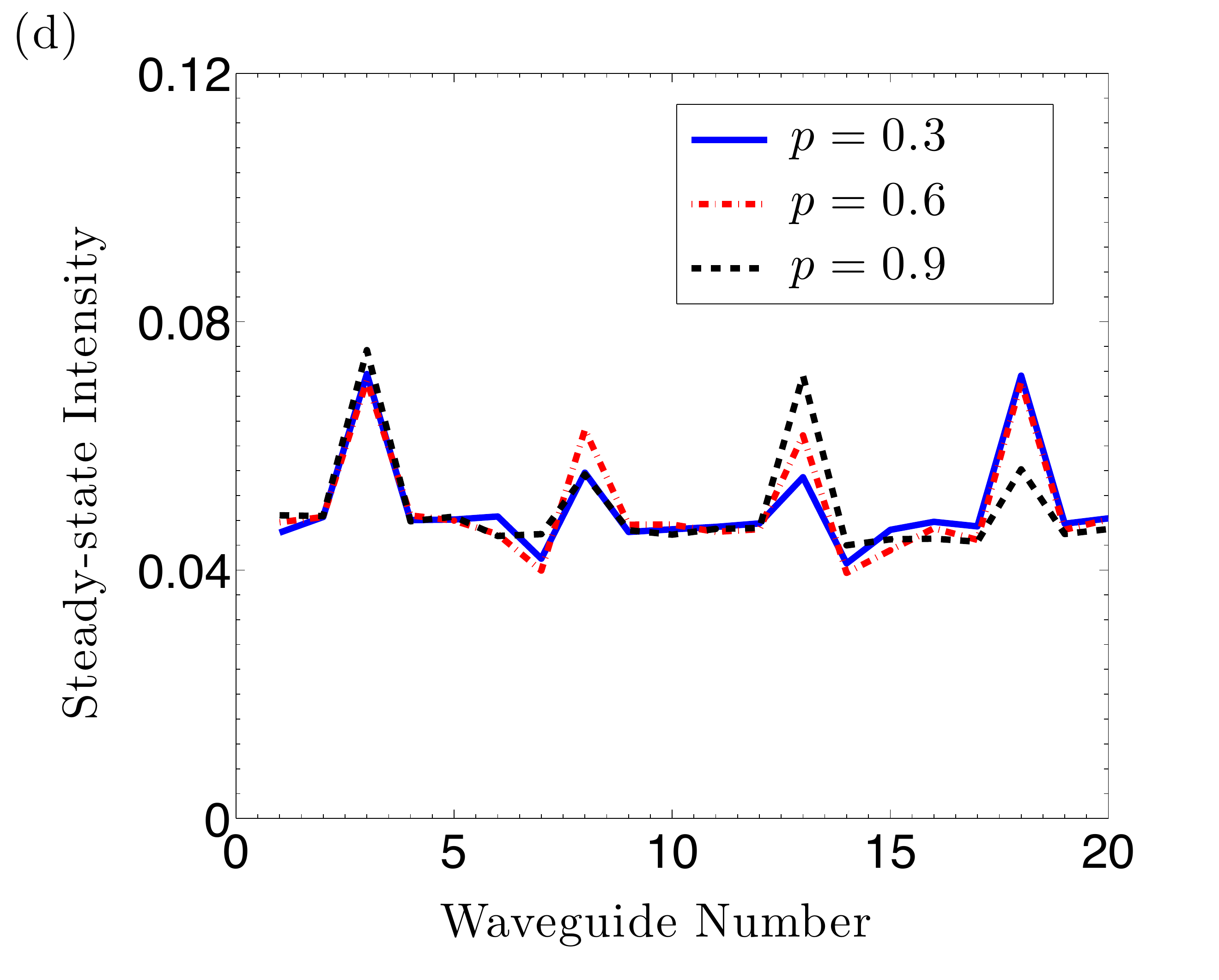}
\caption{(Color Online) Disorder-averaged intensity $\langle I_j(t)\rangle$ in an $N =20$ open, uniform array with weak disorder, $\sigma /C =0.04$ and input light in waveguide $j_0=3$. The boundary Hamiltonian $H_b$ is given by $C(N)/C=p$. Panels (a) $p=0.3$, (b) $p=0.6$, and (c) $p=0.9$ show that the localization peak weight shifts from the parity-symmetric waveguide to the antipodal waveguide as $p$ increases.  Panel (d) shows the steady-state intensity profile for $p=0.3$ (blue solid line), $p=0.6$ (red dashed line), and $p=0.9$ (black dotted line). The intensity at the initial waveguide $j_0=3$ constant. The fractional intensity shifts from its parity-symmetric counterpart $N+1-j_0=18$ to its antipodal counterpart $N/2+j_0=13$ as $p$ increases to one. There is a little intensity transfer to the antipode of the parity-symmetric waveguide, $N/2+1-j_0=8$.}
\label{fig5}
\end{figure*} 
 Lastly, we investigate the crossover of disorder-averaged intensity profile $\langle I_j\rangle$ in small arrays with open boundary conditions to arrays with periodic boundary conditions. The former show localization at the input waveguide and its parity-symmetric counterpart; the latter show localization at the input waveguide and its antipodal counterpart. To model this crossover, we consider a uniform, $N$-waveguide array, $\alpha=0$, with a boundary term $H_b$ given by $C(N)=p C$ where $0\leq p \leq 1$. When $p=0$, the ``boundaries'' of the array are perfectly reflecting, whereas when $p=1$, they are perfectly transmitting. Figures \ref{fig5}(a)-\ref{fig5}(c) show the time evolution of the disorder-averaged intensity $\langle I_j(t)\rangle$ for an initial state at $j_0=3$ in an $N=20$ array with a weak disorder, $\sigma/C=0.04$ and increasing transmission $p$. Panel (a) shows that when the array is ``mostly open'', $p=0.3$, the largest localization peaks occur at the initial waveguide $j_0$ and the parity-symmetric counterpart $N+1-j_0$; in addition, smaller local maxima are also seen at the antipodes of both of these waveguides, $N/2+j_0$ (mod $N$) and $N/2+1-j_0$ (mod $N$). Panels (b) and (c) show that as $p$ increases, the intensities in the parity-symmetric waveguide and its antipodal waveguide are suppressed, and the system moves towards an array with open boundary conditions. Thus, in general, {\it light input into a single waveguide $j_0$ leads to localization peaks at four different waveguides}, namely, $j_0$, $N+1-j_0$, $N/2+j_0$ (mod $N$), and $N/2+1-j_0$ (mod $N$), except some of these indices coincide. 


\section{Boundary-effect quantification}
\label{sec:be}
The results for weak-disorder-induced localization in small waveguide arrays, presented in the previous section, show the importance of interference contribution from (reflected or transmitted) partial waves. This contribution is absent in a truly infinite lattice, and thus ignored in the traditional discussion of localization in waveguide arrays. Experimentally, this is achieved by choosing a disorder that is strong enough so that the time required to develop the steady-state disorder-averaged intensity is shorter than the time required for the partial waves to reach the array boundaries. On the other hand, when the disorder is weak, the localized intensity profile contains unmistakable signatures of these partial waves. In this section, we quantify their contribution. 

For a uniform, infinite array, the time-dependent amplitudes $A(j,t)=\langle j|\psi(t)\rangle$ satisfy the following differential equation,
\be
\label{eq:bessel}
i\partial_t A(j,t)= C\left[ A(j+1,t)+A(j-1,t)\right]. 
\ee
Since the Bessel functions satisfy virtually identical recurrence relation, it follows that the site- and time-dependent amplitudes are given by 
\be
\label{eq:infinite array} 
A_{j_0}(j,t)=i^{j-j_{0}}J_{j-j_{0}}( 2Ct)
\ee
where $j_{0}$ is the location of the initial excitation and $J_n(z)$ is the Bessel function of the first kind~\cite{Jones1965}. We note that the site-dependent phase factor introduces a relative negative sign between amplitudes at next-nearest-neighbor sites. The site- and time-dependent intensity is then given by $I_j(t)=|A(j,t)|^2$. We emphasize that Eq.(\ref{eq:infinite array}) is applicable only for an infinite array and in particular, it is inconsistent with open boundary conditions that require the amplitude to vanish at all times, $A(b,t)=0$ where indices $b$ represent array boundaries.  

It is possible to obtain exact analytical expression for a semi-infinite or finite array using Eq.(\ref{eq:infinite array}) and the method of images~\cite{Makris2006,Szameit2010}. In the presence of a semi-infinite lattice with waveguide indices $n\geq 1$, the time-dependent amplitudes become  
\be 
\label{eq:semi infinite array}
A(j,t)=A_{j_0}(j,t)+ A_{-j_0}(j,t)
\ee
where the second term denotes the contribution from initial input at the ``mirror image'' waveguide and ensures that $A(j,t)$ vanishes identically at the left edge $n=0$. A similar procedure implies that the solution for a finite array with $N$ waveguides is constructed from linear superposition of initial excitations, 
\be
\label{eq:finite array}
A(j,t)=\sum_{r=-\infty}^{\infty} A_{j_0+ 2(N+1)r}(j,t)+ A_{-j_0+2(N+1)r}(j,t),
\ee
located at ``mirror positions'' $j_0+2(N+1)r$ and $-j0+2(N+1)r$. Eq.(\ref{eq:finite array}) ensures that the time-dependent amplitudes vanish identically at $j=0$ and $j=N+1$. In numerical calculations, the infinite sum is truncated at $|r|=R(t)$ where the cutoff $R(t)$ increases linearly with time, since the number of reflections from the edges of the finite array increases linearly with time. 

\begin{figure}[htpb]
\centering
\includegraphics[width=\columnwidth]{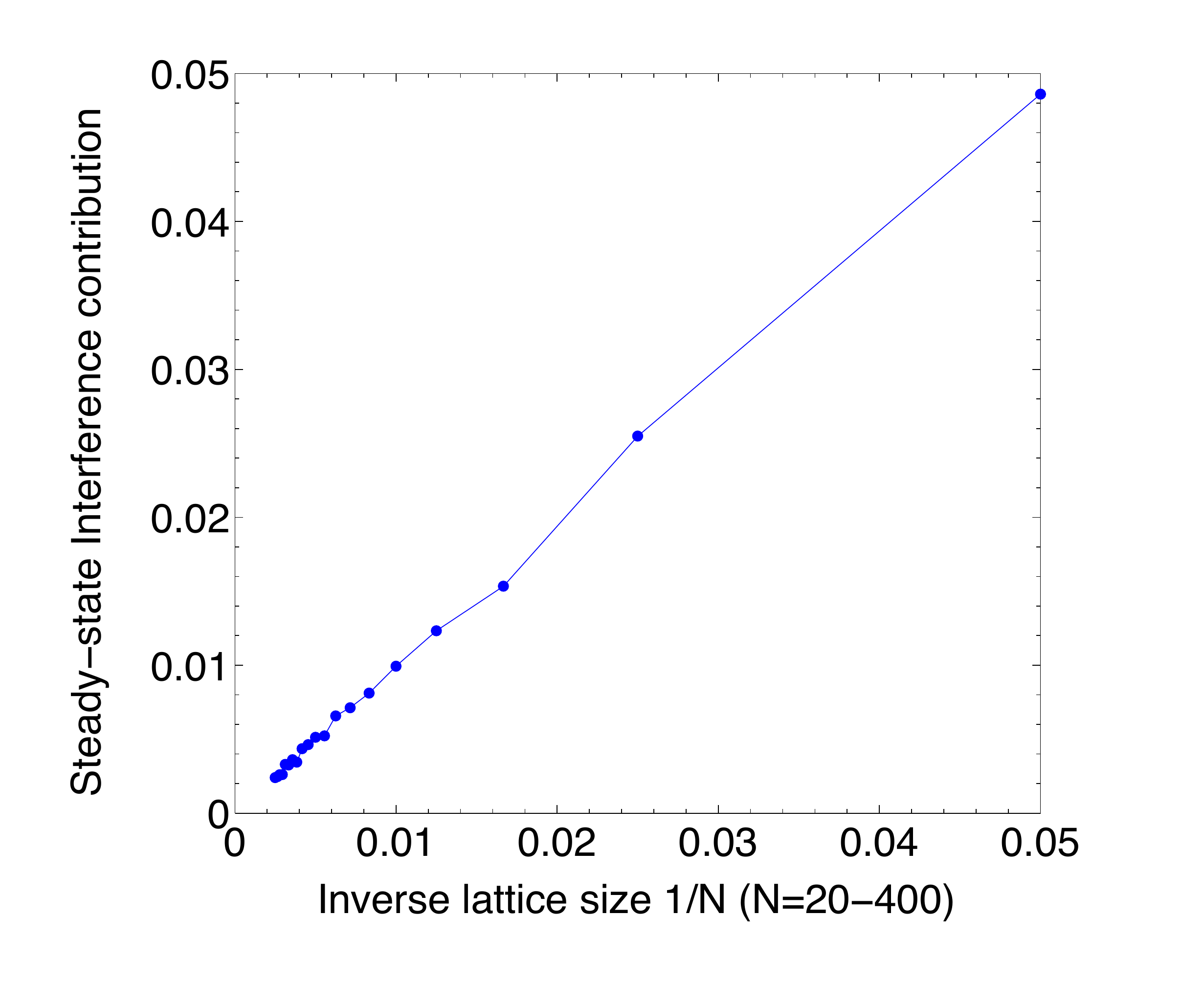}
\caption{(Color Online) Steady-state interference contribution as a function of inverse array-size shows a linear behavior with unit slope, $\mathscr{I}_s(N)=1/N$; these results are independent of the time and initial input waveguide $j_0$, and were obtained by using $R(t)=1500$ image-pair cutoff in Eq.(\ref{eq:intinf}).}
\label{fig3}
\end{figure}
The net interference contribution from the edge reflections is given by adding up all the cross-terms from Eq.(\ref{eq:finite array}) intensity calculation. Thus $\mathscr{I}(t)=\sum_{j=1}^N\mathscr{I}(j,t)$ where 
\begin{eqnarray}
\label{eq:intinf}
\mathscr{I}(j,t)& = & |A(j,t)|^{2}-\sum_{r=-\infty}^{\infty} \left\{|A_{j_0+2(N+1)r}(j,t)|^2\right. \nonumber\\ 
& + & \left.|A_{-j_0+2(N+1)r}(j,t)|^2\right\}.
\end{eqnarray}
The interference contribution $\mathscr{I}(t)$ is zero before the partial waves starting from initial input waveguide $j_0$ reach the boundaries, its maximum value occurs for the first reflection, and it quickly reaches a steady-state value $\mathscr{I}_s$ that is independent of the initial input waveguide $j_0$. Figure~\ref{fig3} shows that the steady-state value, which encodes the effects of multiple reflections, scales linearly with the inverse lattice size. Note that for a weakly disordered array, the light partially localizes in multiple waveguides when $\mathscr{I}_s$ reaches a steady state as the scattering is then able to suppress the disorder-free dynamics. Thus, the multi-waveguide localization that occurs with input in a single waveguide is the result of the competition between scattering due to the disorder and the interference that results from the reflections of the wave packet at the edges of the array.  

\section{Conclusion}
\label{sec:Conclusion}
In this paper, we have examined the effect of boundary conditions on the localization of input light in a small waveguide array in the presence of an extremely weak disorder. In contrast to the well-known exponential localization in an infinite array, we showed that localization in a finite array contains signatures of the array boundary conditions. In particular, we demonstrated that due to boundary reflection or transmission, the input light also localizes in the parity-symmetric waveguide, or the antipodal waveguide, or a combination thereof.  Our results show that ``localization'',  defined as the development of a steady-state site-intensity profile, in small arrays with extremely weak disorder shows a rich behavior that is absent in large arrays with moderate disorder. The dependence of this behavior, including the time to steady-state, the relative fractions of intensity localized to different waveguides, and the generalization to higher dimensions, on system parameters is necessary to fully understand the implications of such localization. 

\bibliography{Sources}{}
\end{document}